\newcommand{\IJQI}[3]{Int.\ Jour.\ Quant. \ Info. {\bf #1},\ #2 (#3)}

\newcommand{\PLA}[3]{Phys.\ Lett.\ A\ {\bf #1},\ #2 (#3)}

\newcommand{\PRL}[3]{Phys.\ Rev.\ Lett.\ {\bf #1},\ #2 (#3)}

\newcommand{\NAT}[3]{Nature\ {\bf #1},\ #2 (#3)}

\newcommand{\PRA}[3]{Phys.\ Rev.\ A\ {\bf #1},\ #2 (#3)}

\newcommand{\JMR}[3]{Jour.\ Mag.\ Res.\ {\bf #1},\ #2 (#3)}

\newcommand{\JPA}[3]{J.\ Phys.\ A:\ Math.\ Gen.\ {\bf #1},\ #2 (#3)}

\newcommand{\AIP}[3]{AIP.\ Conf.\ Proc.\ {\bf #1},\ #2 (#3)}
\newcommand{\NJP}[3]{New.\ Jour.\ Phys.\ {\bf #1},\ #2 (#3)}

\newcommand{\JOSAB}[3]{J.\ Opt.\ Soc.\ Amer.\ B\ {\bf #1},\ #2 (#3)}

\documentclass[aps,showkeys,showpacs,amsmath,amssymb]{revtex4}
\usepackage{dcolumn}
\usepackage{longtable}
\input{Qcircuit}
\begin{document}
\title{Non-destructive discrimination of multiparticle cluster states for quantum computation}
\author{Sreraman Muralidharan}
\affiliation{Loyola College, Chennai - 600034, India}
\email{sreramanm@gmail.com}
\author{Sakshi Jain}
\affiliation{Indian Institute of Technology-Bombay, Mumbai - 400080, India}
\email{sakshi.r.jain@gmail.com}
\author{Sriram Prasath E.}
\email{sridif@gmail.com} \affiliation{Maulana Azad National Institute of Technology, Bhopal - 462051, India}
\author{Prasanta K. Panigrahi}
\email{prasanta@prl.res.in}
\affiliation{Indian Institute of Science Education and Research Kolkata, Mohanpur Campus, BCKV Campus Main Office, Mohanpur - 741252, India}
\affiliation{Physical Research Laboratory, Navrangpura, Ahmedabad - 380 009, India}

\begin{abstract}
Discriminating between orthogonal quantum systems without destroying their entanglement is of interest to quantum computation and communication. In this paper, we explicate the schemes for the non-destructive discrimination (NDD) of 16 orthogonal four qubit cluster states and 32 orthogonal five qubit cluster states respectively. This technique will find its applications in various branches of quantum information theory.
\end{abstract}
\maketitle

\section{Introduction}
Many quantum communication protocols like teleportation \cite{Bennett}, state sharing \cite{Hillery, Qsts} and dense coding \cite{Wiesner} require measurement to be performed on entangled bases. For instance, in the dense coding protocol, involving Bell states as a shared entangled resource, two classical bits are encoded into a qubit by the sender and sent to the receiver. The receiver then performs a measurement on  the Bell basis and recovers the encoded information. In this process, the entanglement is not disturbed and the same Bell state can be reused for the dense coding of a number of classical bits. Many teleportation protocols also involve Bell measurements.
Recently several quantum protocols have been carried out using a special type of graph states known as ``cluster states`` \cite{Robert, Hans} given by
\begin{equation}
|C_N\rangle = \frac{1}{2^{N/2}} \otimes_{a=1}^{N} (|0\rangle_a \sigma_z ^{a+1} + |1\rangle_a), 
\end{equation}
with $\sigma_z^{N+1} = 1$. These states have also been experimentally realized in various systems \cite{cluster1,cluster2,cluster3}. It is well known that they show a strong violation of local reality and are shown to be robust against decoherence \cite{Walther, Hein}. Entanglement exhibited by these states have well been characterized owing to their promising usefulness in Quantum communication protocols \cite{Bai, MEMS}, one way quantum computation \cite{Walther} and also for quantum error correction \cite{Sch}. Two of the present authors \cite{Sre2} have shown them to be useful for perfect teleportation and state sharing of an arbitrary two qubit state. Its superdense coding capacity reaches the Holevo bound. The main hurdle towards the experimental realization of the above mentioned protocols is that most of these schemes involve arbitrary multiqubit measurements \cite{Sre1, Sre2, Sre3, Yeo, Sre4} which are extremely difficult to realize in laboratory conditions. 

In an experimental scenario, since even performing a simple Bell basis measurement is a non trivial task, measurement on a product basis is preferred. Recently, this has been made possible by performing suitable quantum gates which are experimentally realizable. Indeed, this technique has been used for the experimental implementation of teleportation in $NMR$ \cite{Nielson2} and in quantum dots \cite{qdot} for the teleportation of an arbitrary single qubit state. However, the disadvantage arising out of this technique is that the same resource state cannot be reused for other quantum protocols since the entanglement is destroyed. 

	Here, we propose an appropriate circuit using which we perform a non-destructive measurement of an entangled basis. Recently, it was shown using an appropriate quantum circuit that four Bell pairs can be discriminated       deterministically \cite{Manu1, Manu2}. The advantage of our scheme over the previous schemes discussed above is that we perform a product basis measurement and yet do not disturb the entanglement of the system. We use this method to discriminate between orthogonal four and five qubit cluster states.
	
The paper is organized as follows:	We initially discuss the circuits for the physical realization of four and five qubit cluster states. The next section deals with the explicit
 circuits for their NDD. In the last section, we illustrate the usefulness of our circuit for various quantum
 computational protocols.
\section{Physical Realization} 
	Explicit circuits for the physical realization of the multi-particle cluster states have been studied by two of the present authors \cite{Sre5}. For the sake of completeness, we describe these circuits before we proceed with the non-destructive discrimination
of these states. The 16 orthogonal four qubit cluster states can be generated by implementing the following general circuit diagram (Fig. 1) 
			\begin{figure}[h]
	\caption{Circuit diagram for the generation of $|C_4\rangle$}
	\label{fig:CircuitDiagram}
	\leavevmode
\centering	
\Qcircuit @C=3em @R=2em {
\lstick{\ket{0}}& \gate{H} &\ctrlo{1} & \qw      & \ctrl{2}               & \qw           \\
\lstick{\ket{0}}& \qw &\targ     & \qswap   & \ctrl{-1}              & \qw     & |C_4\rangle   \\
\lstick{\ket{0}}& \gate{H} &\ctrlo{1} & \qw \qwx   &  \gate{Z}      & \qw    \\
\lstick{\ket{0}}& \qw &\targ     & \qswap \qwx   &       \qw  		  & \qw    
\ \gategroup{1}{1}{4}{6}{.7em}{\}} }  \\

\end{figure}
by changing the inputs $|0000\rangle_{1234}$ to a different computational bases.
Similarly, the five-qubit cluster states can be generated by the following circuit (Fig. 2) 
				\begin{figure}[h]
	\caption{Circuit diagram for the generation of $|C_5\rangle$}
	\label{fig:CircuitDiagram}
	\leavevmode
\centering	
\Qcircuit @C=3em @R=2em {
\lstick{\ket{0}}& \gate{H} &\ctrlo{1} & \qw      & \ctrlo{2}               & \qw           & \qw \\
\lstick{\ket{0}}& \qw &\targ     & \ctrlo{1}    & \qw \qwx              & \qw       & \qw \\
\lstick{\ket{0}}& \gate{H} &\ctrlo{1} & \targ \qwx    &  \qw \qwx      & \qw    & \qw & |C_5\rangle\\
\lstick{\ket{0}}& \qw &\targ     &\ctrlo{1} \qw   &     \qw   \qwx  		  & \qw      & \qw \\
\lstick{\ket{0}}& \qw &\qw     & \targ \qwx   & \targ \qwx              & \qw       & \qw 
\ \gategroup{1}{1}{5}{7}{.7em}{\}} }  \\

\end{figure}
by changing the inputs $|00000\rangle_{12345}$ to a different computational bases.
Now, we shall proceed with the non-destructive discrimination of these states.

\section{Non-Destructive Discrimination}
\subsection{four-qubit Cluster States}
The sixteen orthogonal cluster states can be distinguished without disturbing the system by carrying out measurements on four ancilla bits. The four qubit cluster states are made to interact with these ancillas as shown in Fig 3. The outcome of the measurement on the ancillas reveals the respective cluster state as shown in the Table-I:
				
\begin{figure}[h]
	\caption{Circuit diagram for NDD of $|C_4\rangle$}
	\label{fig:CircuitDiagram}
	\leavevmode
\centering	
\Qcircuit @C=0.5em @R=0.4em @! {
\lstick{} & \qw & \ctrl{1} & \ctrl{1} & \qw & \targ & \qw & \qw& \qw & \qw& \qw& \qw & \ctrl{1} & \qw\\
\lstick{|C_4\rangle} & \qw & \qw \qwx & \qw \qwx & \ctrl{1} & \qw \qwx & \targ & \qw& \qw& \qw& \qw& \qw& \qw \qwx& \qw\\
\lstick{} & \qw & \ctrl{1} \qwx & \qw \qwx & \qw \qwx & \qw \qwx  & \qw \qwx & \ctrl{1}& \qw& \targ& \qw& \qw& \ctrl{1} \qwx& \qw & |C_4\rangle\\
\lstick{} & \qw &  \gate{Z} &  \qw \qwx & \qw \qwx & \qw \qwx & \qw \qwx & \qw \qwx & \ctrl{1} & \qw \qwx& \targ& \qw&  \gate{Z}& \qw\\
\lstick{\ket{0}} & \qw & \qw & \targ \qwx & \targ \qwx & \qw \qwx & \qw \qwx & \qw \qwx& \qwx \qw& \qw\qwx & \qw \qwx& \qw& \meter\\
\lstick{\ket{0}} & \qw & \gate{H} & \qw & \qw & \ctrl{-1}  & \ctrl{-1} & \qw \qwx & \qwx \qw& \qw\qwx & \qw \qwx& \gate{H}& \meter\\
\lstick{\ket{0}} & \qw & \qw & \qw & \qw & \qw & \qw& \targ \qwx& \targ \qwx& \qw\qwx & \qw \qwx& \qw & \meter\\
\lstick{\ket{0}} & \qw & \gate{H} & \qw & \qw & \qw & \qw& \qw& \qw& \ctrl{-1}& \ctrl{-1}& \gate{H}& \meter
\ \gategroup{1}{14}{4}{14}{.7em}{\}}  \gategroup{1}{1}{4}{1}{.7em}{\{} }  \\
\end{figure}

				\begin{table}[h]
\caption{\label{tab5}Four-qubit Cluster State discrimination from the ancilla measurements}
\begin{tabular}{|c|c|}
\hline {\bf Ancilla Measurement}& {\bf Corresponding four-qubit Cluster State}\\ 
\hline

$|0000\rangle$&$|0000\rangle + |0011\rangle + |1100\rangle -|1111\rangle$\\
$|0001\rangle$&$|0000\rangle - |0011\rangle + |1100\rangle +|1111\rangle$\\
$|0010\rangle$&$|0001\rangle + |0010\rangle - |1101\rangle +|1110\rangle$\\
$|0011\rangle$&$|0001\rangle - |0010\rangle - |1101\rangle -|1110\rangle$\\
$|0100\rangle$&$|0000\rangle + |0011\rangle - |1100\rangle +|1111\rangle$\\
$|0101\rangle$&$|0000\rangle - |0011\rangle - |1100\rangle -|1111\rangle$\\
$|0110\rangle$&$|0001\rangle + |0010\rangle + |1101\rangle -|1110\rangle$\\
$|0111\rangle$&$|0001\rangle - |0010\rangle + |1101\rangle +|1110\rangle$\\
$|1000\rangle$&$|0100\rangle + |0111\rangle + |1000\rangle -|1011\rangle$\\
$|1001\rangle$&$|0100\rangle - |0111\rangle + |1000\rangle +|1011\rangle$\\
$|1010\rangle$&$|0101\rangle + |0110\rangle - |1001\rangle +|1010\rangle$\\
$|1011\rangle$&$|0101\rangle - |0110\rangle - |1001\rangle -|1010\rangle$\\
$|1100\rangle$&$|0100\rangle + |0111\rangle - |1000\rangle +|1011\rangle$\\
$|1101\rangle$&$|0100\rangle - |0111\rangle - |1000\rangle -|1011\rangle$\\
$|1110\rangle$&$|0101\rangle + |0110\rangle + |1001\rangle -|1010\rangle$\\
$|1111\rangle$&$|0101\rangle - |0110\rangle + |1001\rangle +|1010\rangle$\\
 \hline
\end{tabular}
\end{table}

Since our scheme involves only CNOT \cite{Cnot}, Hadamard \cite{Hada} and controlled phase shift gates \cite{Cphase}, it is completely feasible.

\subsection{NDD of five qubit cluster states}
There are 32 orthogonal $|C_{5}\rangle$ states which need to be distinguished for several quantum protocols. For this purpose, we make use of the five ancillas and make it interact with the five qubit cluster state 
as per Fig. 4. The outcome of the measurement on the ancillas and the corresponding cluster state are shown in the Table-II.

\begin{figure}[h]
	\caption{Circuit diagram for NDD of $|C_5\rangle$}
	\label{fig:CircuitDiagram}
	\leavevmode
\centering	
\Qcircuit @C=0.5em @R=0.4em @! {
\lstick{} & \qw &\ctrl{1} & \qw & \ctrl{1} & \qw & \targ & \qw & \qw& \qw & \qw& \qw& \qw & \qw & \ctrl{1}& \qw& \qw\\
\lstick{|C_5\rangle} &\qw & \qw \qwx & \qw  & \qw \qwx & \ctrl{1} & \qw \qwx & \targ & \qw& \qw& \qw& \qw& \qw& \qw & \qw\qwx& \qw& \qw\\
\lstick{} & \ctrl{1} &\targ \qwx & \qw  & \qw \qwx & \qw \qwx & \qw \qwx  & \qw \qwx & \ctrl{1}& \qw& \targ& \qw& \qw& \qw &  \targ \qwx& \ctrl{1} & \qw & |C_5\rangle\\
\lstick{} & \qw \qwx &\qw &  \qw &  \qw \qwx & \qw \qwx & \qw \qwx & \qw \qwx & \qw \qwx & \ctrl{1} & \qw \qwx& \targ& \qw& \qw& \qw& \qw\qwx& \qw\\
\lstick{} & \targ \qwx &\qw &  \qw &  \qwx \qw  & \qw \qwx  & \qw \qwx & \qw  \qwx & \qw  \qwx & \qw \qwx & \qw \qwx & \qw \qwx&  \qw& \qw & \ctrl{1}&  \targ \qwx & \qw\\
\lstick{\ket{0}} & \qw &\qw & \qw & \targ \qwx & \targ \qwx & \qw \qwx & \qw \qwx & \qw \qwx& \qwx \qw& \qw\qwx & \qw \qwx& \qw& \qw& \qw \qwx&  \meter\\
\lstick{\ket{0}} & \qw &\qw  & \gate{H} & \qw & \qw & \ctrl{-1}  & \ctrl{-1} & \qw \qwx & \qwx \qw& \qw\qwx & \qw \qwx& \gate{H}& \qw& \qw \qwx & \meter\\
\lstick{\ket{0}} & \qw &\qw & \qw & \qw & \qw & \qw & \qw& \targ \qwx& \targ \qwx& \qw\qwx & \qw \qwx& \qw &\qw& \qw\qwx & \meter\\
\lstick{\ket{0}} & \qw &\qw & \gate{H} & \qw & \qw & \qw & \qw& \qw& \qw& \ctrl{-1}& \ctrl{-1}& \gate{H}&\qw& \qw \qwx & \meter\\
\lstick{\ket{0}} & \qw & \qw & \qw &  \qw  & \qw  & \qw  & \qw  & \qw  & \qw & \qw & \qw& \qw&  \qw& \targ \qwx & \meter\\
 \ \gategroup{1}{17}{5}{17}{.7em}{\}}  \gategroup{1}{1}{5}{1}{.7em}{\{} }  \\
  
\end{figure}
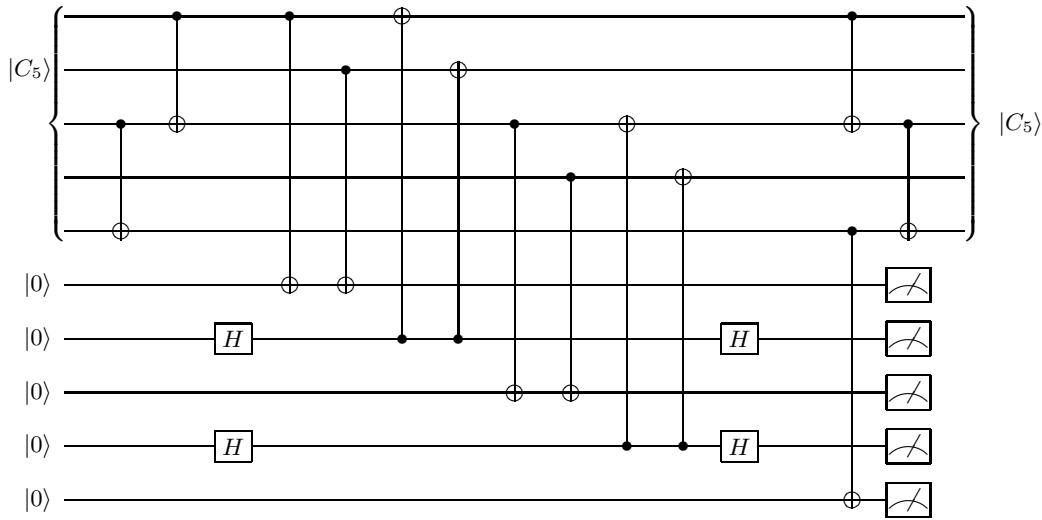

\begin{table}[h]
\caption{\label{tab4}Five qubit cluster state discrimination from the ancilla measurements}
\begin{tabular}{|c|c|}
\hline {\bf Ancilla Measurement}& {\bf Corresponding five qubit Cluster State}\\ 
\hline

$|00000\rangle$&$|00000\rangle + |00111\rangle + |11101\rangle +|11010\rangle$\\
$|00001\rangle$&$|00000\rangle - |00111\rangle + |11101\rangle -|11010\rangle$\\
$|00010\rangle$&$|00010\rangle + |00101\rangle + |11111\rangle -|11000\rangle$\\
$|00011\rangle$&$|00010\rangle - |00101\rangle + |11111\rangle -|11000\rangle$\\
$|00100\rangle$&$|00000\rangle + |00111\rangle - |11101\rangle -|11010\rangle$\\
$|00101\rangle$&$|00000\rangle - |00111\rangle - |11101\rangle +|11010\rangle$\\
$|00110\rangle$&$|00010\rangle + |00101\rangle - |11111\rangle -|11000\rangle$\\
$|00111\rangle$&$|00010\rangle - |00101\rangle - |11111\rangle +|11000\rangle$\\
$|01000\rangle$&$|01000\rangle + |01111\rangle + |10101\rangle +|10010\rangle$\\
$|01001\rangle$&$|01000\rangle - |01111\rangle + |10101\rangle -|10010\rangle$\\
$|01010\rangle$&$|01010\rangle + |01101\rangle + |10111\rangle +|10000\rangle$\\
$|01011\rangle$&$|01010\rangle - |01101\rangle + |10111\rangle -|10000\rangle$\\
$|01100\rangle$&$|01000\rangle + |01111\rangle - |10101\rangle -|10010\rangle$\\
$|01101\rangle$&$|01000\rangle - |01111\rangle - |10101\rangle +|10010\rangle$\\
$|01110\rangle$&$|01101\rangle + |01010\rangle - |10000\rangle -|10111\rangle$\\
$|01111\rangle$&$|01010\rangle - |01101\rangle - |10111\rangle +|10000\rangle$\\
$|10000\rangle$&$|00001\rangle + |00110\rangle + |11100\rangle +|11011\rangle$\\
$|10001\rangle$&$|00001\rangle - |00110\rangle + |11100\rangle -|11011\rangle$\\
$|10010\rangle$&$|00011\rangle + |00100\rangle + |11110\rangle +|11001\rangle$\\
$|10011\rangle$&$|00011\rangle - |00100\rangle + |11110\rangle -|11001\rangle$\\
$|10100\rangle$&$|00001\rangle + |00110\rangle - |11100\rangle -|11011\rangle$\\
$|10101\rangle$&$|00001\rangle - |00110\rangle - |11100\rangle +|11011\rangle$\\
$|10110\rangle$&$|00011\rangle + |00100\rangle - |11110\rangle -|11001\rangle$\\
$|10111\rangle$&$|00011\rangle - |00100\rangle - |11110\rangle +|11001\rangle$\\
$|11000\rangle$&$|01001\rangle + |01110\rangle + |10100\rangle +|10011\rangle$\\
$|11001\rangle$&$|01001\rangle - |01110\rangle + |10100\rangle -|10011\rangle$\\
$|11010\rangle$&$|01011\rangle + |01100\rangle + |10110\rangle +|10001\rangle$\\
$|11011\rangle$&$|01011\rangle - |01100\rangle + |10110\rangle -|10001\rangle$\\
$|11100\rangle$&$|01001\rangle + |01110\rangle - |10100\rangle -|10011\rangle$\\
$|11101\rangle$&$|01001\rangle - |01110\rangle - |10100\rangle +|10011\rangle$\\
$|11110\rangle$&$|01011\rangle + |01100\rangle - |10110\rangle -|10001\rangle$\\
$|11111\rangle$&$|01011\rangle - |01100\rangle - |10110\rangle +|10001\rangle$\\ \hline
\end{tabular}
\end{table}

\section{Illustration}
\subsection{Quantum dialogue using NDD of cluster states}
We now devise a scheme for "quantum dialogue" using reusable cluster states with dense coding. 
The four and the five qubit cluster states reach the "Holevo bound" 
allowing four classical bits to be transmitted by sending two qubits and five classical
bits be transmitted by sending three qubits respectively. In the standard dense coding protocol \cite{Wiesner} involving
four qubit cluster states, Alice and Bob possess the first and the last two qubits respectively. 
Alice encodes four classical bits of information into two qubits by performing 
a ($\sigma_{i} \otimes \sigma_{j}$) operation where $i,j \in (0,1,2,3)$.
Alice, now sends her qubit to Bob. On receiving the qubits from Alice, Bob
can perform our NDD scheme to decode the message without destroying $|C_4\rangle$.
This is shown in Fig. 5.
 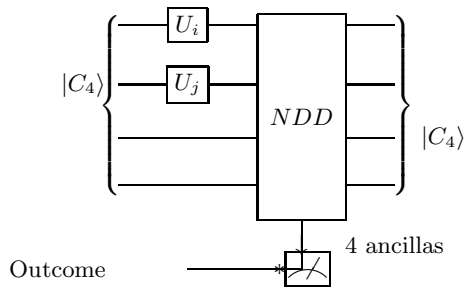
\begin{figure}[h]
	\caption{Circuit diagram for quantum dialogue using NDD}
	\label{fig:CircuitDiagram}
	\leavevmode
\centering	
\Qcircuit @C=2em @R=1em {
\lstick{}& \gate{U_i}     &\multigate {4}{NDD}   &\qw  \\
\lstick{\ket{C_4}}& \gate{U_j}     &\ghost{NDD} & \qw               \\
\lstick{}& \qw            &\ghost{NDD} & \qw               & |C_4\rangle               \\
\lstick{}& \qw            &\ghost{NDD} & \qw                  \\
\lstick{}&             &&       &                &           &  \\
\lstick{}&             & \qwx \downarrow&\text{4 ancillas}      &                &           &  \\
\lstick{\text{Outcome}}&   &  \qwx  \meter &       &                &           &  \\
\ \gategroup{1}{1}{4}{1}{.7em}{\{}  \gategroup{1}{4}{4}{4}{.7em}{\}} 
}  \\
 \end{figure}

Now, Bob can send two qubits of $|C_4\rangle$ to Alice and
 encode four cbits of information  by operating
on her two qubits and send them to Alice. Alice can decode this 
information through NDD and the dialogue continues without
destroying the entanglement of the shared channel.
 We can construct a similar quantum circuit for this scheme using $|C_5\rangle$
in which case Alice can encode five cbits in three qubits. 
\subsection{Error detection}
The three prominent types of error that can occur in a quantum system 
are "bit flip" error,  "phase flip" error or both. These errors
yield sixteen different orthogonal states. Since our NDD scheme
helps in discriminating these orthogonal states, it is possible
for one to know the position of the error that has occured in the
intial cluster state. This doesnot require one to perform a cluster
basis measurement to know the position and the type of the error.
After the position and the type of the error are known, one can 
correct it by using appropriate quantum gates.  
\section{Conclusion}
In this paper, we presented a scheme for discrimination of orthogonal multipartite
cluster states without destroying the state. We described explicit circuit
diagrams to achieve the same. This circuit can be attached to any quantum 
protocol requiring cluster basis measurement for preserving the entanglement
of the resource. We also illustrated the utility of our schemes by devising
a quantum dialogue protocol involving cluster states. We hope that this method
will find its application in other branches of quantum information that require
cluster states as an initial resource. This technique is alo experimentally
feasible as it requires measurement to be performed only on a product basis.
In future, we wish to study in detail the usefulneess of these schemes for other quantum protocols. 

\begin{acknowledgments}
 The work was supported by the summer project programme undertaken by the Indian Institute of Science Education and Research (Kolkata), India. The authors acknowledge Sidharth K for discussions.

\end{acknowledgments}

\end{document}